\documentclass[%
 aip,jcp,
 amsmath,amssymb,
 preprint,%
 a4paper]{revtex4-1}

\usepackage{graphicx}
\usepackage{dcolumn}
\usepackage{bm}

\usepackage[utf8]{inputenc}
\usepackage[T1]{fontenc}
\usepackage{mathptmx}
\usepackage{etoolbox}

\usepackage{xcolor}
\usepackage[version=4]{mhchem}
\usepackage{units}

\makeatletter
\def\@email#1#2{%
 \endgroup
 \patchcmd{\titleblock@produce}
  {\frontmatter@RRAPformat}
  {\frontmatter@RRAPformat{\produce@RRAP{*#1\href{mailto:#2}{#2}}}\frontmatter@RRAPformat}
  {}{}
}%
\makeatother

\draft 

\begin{document}


\title[Auger Decay of Double Core Holes]{Radiationless Decay Spectrum of O~1s Double Core Holes in Liquid Water}


\author{Florian~Trinter}
\email{trinter@fhi-berlin.mpg.de}
\affiliation{Fritz-Haber-Institut der Max-Planck-Gesellschaft, Faradayweg~4-6, 14195 Berlin, Germany}
\affiliation{Institut für Kernphysik, Goethe-Universität Frankfurt, Max-von-Laue-Str. 1, 60438 Frankfurt am Main, Germany}

\author{Ludger~Inhester}
\email{ludger.inhester@cfel.de}
\affiliation{Center for Free-Electron Laser Science CFEL, Deutsches Elektronen-Synchrotron DESY, Notkestr. 85, 22607 Hamburg, Germany}

\author{Ralph~Püttner}
\affiliation{Fachbereich Physik, Freie Universität Berlin, Arnimallee~14, 14195 Berlin, Germany}

\author{Sebastian~Malerz}
\affiliation{Fritz-Haber-Institut der Max-Planck-Gesellschaft, Faradayweg~4-6, 14195 Berlin, Germany}

\author{Stephan~Thürmer}
\affiliation{Department of Chemistry, Graduate School of Science, Kyoto University, Kitashirakawa-Oiwakecho, Sakyo-Ku, 606-8502 Kyoto, Japan}

\author{Tatiana~Marchenko}
\affiliation{Sorbonne Université, CNRS, Laboratoire de Chimie Physique -- Matière et Rayonnement, LCPMR, 75005 Paris, France}

\author{Maria~Novella~Piancastelli}
\affiliation{Sorbonne Université, CNRS, Laboratoire de Chimie Physique -- Matière et Rayonnement, LCPMR, 75005 Paris, France}

\author{Marc~Simon}
\affiliation{Sorbonne Université, CNRS, Laboratoire de Chimie Physique -- Matière et Rayonnement, LCPMR, 75005 Paris, France}

\author{Bernd~Winter}
\affiliation{Fritz-Haber-Institut der Max-Planck-Gesellschaft, Faradayweg~4-6, 14195 Berlin, Germany}

\author{Uwe~Hergenhahn}
\email{hergenhahn@fhi-berlin.mpg.de}
\affiliation{Fritz-Haber-Institut der Max-Planck-Gesellschaft, Faradayweg~4-6, 14195 Berlin, Germany}

\date{\today}

\begin{abstract}
We present a combined experimental and theoretical investigation of the radiationless decay spectrum of an O~1s double core hole in liquid water.
Our experiments were carried out using liquid-jet electron spectroscopy from cylindrical microjets of normal and deuterated water.
The signal of the double-core-hole spectral fingerprints (hypersatellites) of liquid water is clearly identified, 
with an intensity ratio to Auger decay of singly charged O~1s of 0.0014(5).
We observe a significant isotope effect between liquid H$_2$O and D$_2$O.
For theoretical modeling, the Auger electron spectrum of the central water molecule in a water pentamer was calculated using an electronic-structure toolkit combined with molecular-dynamics simulations to capture the influence of molecular rearrangement on the ultrashort lifetime of the double core hole.
We obtained the static and dynamic Auger spectra for H$_2$O, (H$_2$O)$_5$, D$_2$O, and (D$_2$O)$_5$, instantaneous Auger spectra at selected times after core-level ionization, and the symmetrized oxygen-hydrogen distance as a function of time after double core ionization for all four prototypical systems.
We consider this observation of liquid-water double core holes as a new tool to study ultrafast nuclear dynamics.
\end{abstract}


\maketitle 

\section{Introduction}
%
As nature's most important liquid, water remains of continuing interest to natural scientists because of its key role in chemistry and biology, since most chemical reactions and biological functions occur in aqueous environments, and because of its anomalous properties.\cite{Pettersson2016}
The latter are intimately linked to the hydrogen-bonding structure in liquid water.
Various X-ray spectroscopy techniques, sensitive to the chemical environment of a specific element, have been shown to be efficient probes of water's electronic and hydrogen-bond structure.\cite{Fransson2016}
Photoemission spectra recorded upon ionization in the soft-X-ray region reflect the electronic and nuclear dynamics in aqueous solution (see, e.g., Refs.~\citenum{Winter2005, Thuermer2013, Unger2015, Slavicek2016, Hollas2017, Saak2020}).
The simultaneous presence of electronic and nuclear dynamics in liquid water complicates the interpretation of the measurements.
In a number of studies, altering the nuclear dynamics via isotopic substitution was used to aid in separating these two contributions.\cite{Brena2004, Thuermer2013, Slavicek2014}
The dynamics in water induced by core-level ionization has also been studied by X-ray emission spectroscopy of oxygen $K$-shell vacancies, by exploring the excitation-energy dependence of the spectrum and the isotope effect.\cite{Odelius2005, Fuchs2008, Odelius2009, Ertan2018}
Another aspect of the interaction of water with ionizing radiation is the production of reactive decay products, such as OH radicals, ions, and free electrons, generated by the primary ionization followed by a cascade of reactions and secondary ionization processes.\cite{Alizadeh2012, Stumpf2016, Gopakumar2023}
Therefore, knowledge of the dynamic response of water to ionizing X-rays is an important aim of current water studies.

The time scale for the decay of a single oxygen core hole in liquid water can be estimated from the lifetime broadening in the O~1s main line of gaseous water as 4~fs.\cite{Sankari}
A window into even shorter time scales is provided by the decay of double core holes (DCHs).\cite{Inhester2012}
This refers to a state with two inner-shell holes created in sufficient temporal and spatial proximity to interact with each other.
Whereas for single-core-level ionization, the liquid environment around a water molecule substantially influences the ultrafast core-hole-induced proton dynamics,\cite{Thuermer2013, Slavicek2014} the impact of the chemical environment, i.e., substances in the close neighborhood of the molecule, has so far not been addressed for DCHs.
In an isolated water molecule, the DCH leads to an ultrafast dissociation of both hydrogen atoms as protons.\cite{Inhester2012, Marchenko2018}
It remains an open question how strong the chemical environment influences this ultrafast dissociation.
Moreover, for a single core hole, neighboring water molecules can to some extent directly participate in the core-hole decay via intermolecular Coulombic decay or via electron-transfer-mediated decay processes,\cite{Thuermer2013, Slavicek2014} which have not been reported for DCHs.
Here, we will present experimental and theoretical results on the Auger decay of DCHs produced by simultaneously ionizing both oxygen core electrons of one water molecule in the liquid phase.

Double core holes were first discovered in the study of $K$-shell capture in radioactive isotopes, which were found to contain two $K$-shell holes in a single metal atom for some decay pathways.\cite{Charpak1953, Perlman1954}
Analogous to single core holes they may relax by either Auger decay or X-ray emission.
The latter became the preferred probe to observe double $K$-shell holes in the decades following their first observation.
Significant energy shifts in X-ray emission of $K$-shell DCHs, compared to the characteristic X-rays of the same element, were observed which led to the term 'hypersatellites' for these X-ray lines.\cite{Briand1971}
Studies of the Auger decay of DCHs greatly accelerated after it became apparent that these states are also created in ion-atom collisions; the term hypersatellites soon was also used to designate the energy-shifted Auger lines emerging from DCHs.\cite{Woods1975}

\begin{figure}
\includegraphics[width=0.7\columnwidth]{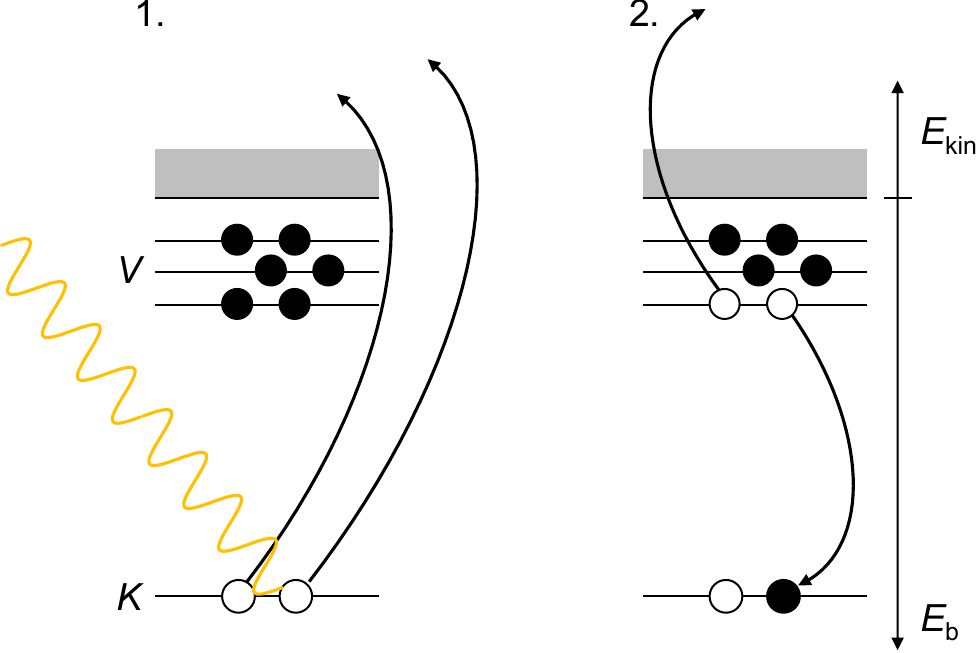}%
\caption{
Schematic showing the simultaneous ejection of two core electrons (left, label 'K') and the first step in the radiationless decay of the two core holes (right) by rearrangement of the valence shell (label 'V').
The kinetic energy ($E_{\rm kin}$) spectrum of the latter decay step in liquid water will be discussed here.
$E_{\rm b}$ denotes the binding energy.
\label{fig:scheme}}%
\end{figure}

In recent years, strong experimental and theoretical activities have been devoted to the study of inner-shell double photoionization,\cite{Penent2015} prompted both by the perspective to create such states using X-ray free-electron lasers (XFELs) with very high brightness,\cite{Cryan2010, Fang2010} as well as by the availability of sophisticated coincidence detection schemes that greatly enhance the amount of information on DCHs compared to using more conventional synchrotron light sources (see, e.g., Refs.~\citenum{Eland2010, Lablanquie2011, Lablanquie2011a, Linusson2011, Mucke2013, Nakano2013, Carniato2015, Carniato2015a, Lablanquie2016}).
The high brightness of XFELs has promoted successive absorption of many X-ray photons by atoms at the atomic inner shells within few tens of femtoseconds, leading to high spectral complexity.\cite{Schaefer2018}
The ensuing decay products could nevertheless be disentangled by suitable analysis methods.\cite{Frasinski2013}
We do not touch upon this type of experiments here further, but restrict ourselves to DCHs produced by absorption of a single photon from a synchrotron radiation source.
The DCH creation and an exemplary radiationless decay channel are sketched in Fig.~\ref{fig:scheme}.
Mechanisms for single-photon double photoionization in general were delineated and compared, e.g., in Ref.~\citenum{Rost2002}.
These authors distinguish between {knockout} -- essentially an electron-electron collision -- and {shake-off}, a quantum mechanical effect where the non-orthogonality of initial-state orbitals with a continuum wave function in the singly photoionized state fosters the transition of a second, bound electron into the continuum.
For DCHs a number of studies have targeted the latter of these mechanisms, but mostly in the context of singly ionized states that feature an additional core-to-valence {\em excitation} (as opposed to {\em ionization}, in our experiment); see Refs.~\citenum{Nakano2013, Ferte2020, Tenorio2021} and with a focus on water Ref.~\citenum{Carniato2015}.
A complete picture of the creation of water DCHs is missing so far, however.

In our experiment, DCHs are created at a single molecular site of liquid water, i.e., we discuss states with configuration \ce{(H2O^2+ {1s}^{-2})(H2O)_N(aq)}.
This doubly ionized state is expected to relax via several radiationless decay steps.
In gaseous water, the hypersatellite Auger spectrum of the first decay step was investigated experimentally and theoretically.\cite{Inhester2012, Inhester2012err, Marchenko2018, Ismail2023}
The first radiationless decay step was found to take place within 1.5~fs (compared to 4~fs lifetime of a singly ionized core hole in water, see above). 
In this sense, the liquid-water DCH can be considered as a new tool for studying ultrafast nuclear dynamics on a yet shorter time scale.
Here, we will present the electronic decay spectra of the same vacancy states measured from liquid water.
In an attempt to isolate the effects of nuclear dynamics, measurements will be presented for normal and heavy water.
Our experiment opens a window into liquid-water nuclear dynamics in the one-femtosecond range (i.e., faster compared to previous single-core-hole studies in solution), as we do not assume that the DCH lifetime is strongly influenced by the liquid environment.

For gaseous water the experimental results showed that besides 1s$^{-2}$ DCHs also the decay of satellite states,\cite{Marchenko2018, Tenorio2021, ismail2024ultrafast} featuring a 1s$^{-2}$ DCH and an additional, excited electron-hole pair, plays a role.
A greater importance of satellite states compared to single core ionization is common for DCHs; this leads not only to the creation of DCH states with an additional valence excitation, but also to singly ionized states featuring core-hole ionization plus excitation of a second core electron into the valence shell.\cite{Carniato2020, Ferte2020, Ferte2022}
The latter states constitute a series of resonances energetically located around the DCH production threshold.
They are reached by ejection of a single electron, while creation of a non-resonant DCH state requires ejection of an electron pair.
The primary electrons pertaining to the resonant DCH satellites therefore can be detected by conventional electron spectroscopy, for experiments in gaseous water see Refs.~\citenum{Carniato2015, Marchenko2020}.
For liquid water we will return to this aspect below.
The pair of primary electrons giving rise to non-resonant DCHs and doubly charged DCH satellite states can arbitrarily share the available excess energy, and can only be identified by coincidence-detection methods that were not available for this study.\cite{Eland2010, Lablanquie2011, Lablanquie2011a, Linusson2011, Mucke2013}

Several recent publications have discussed the hypersatellite Auger spectra of DCHs in other gaseous systems.
Detailed results appeared on Ne, isoelectronic to water.\cite{Goldsztejn2016, Goldsztejn2017}
For a molecular sample (\ce{CH3I}), the potential energy landscape of a DCH state could be extracted from high-resolution data.\cite{Marchenko2017}
The strongly repulsive nature of these states, which leads to ultrafast dissociation, was illustrated in Refs.~\citenum{Travnikova2016, Travnikova2017} and will be discussed for water in this work.

\section{Methods}

\subsection{Experimental Methods}
The liquid-water experiments were performed using the EASI (Electronic structure from Aqueous Solutions and Interfaces) liquid-jet photoelectron spectroscopy instrument,\cite{EASI} installed at the soft-X-ray beamline P04 (Ref.~\citenum{Viefhaus2013}) of the synchrotron radiation source PETRA~III (DESY, Hamburg, Germany).
Data acquired in four different campaigns have been compiled for this work.
EASI is based on a Scienta-Omicron HiPP3 differentially pumped hemispherical electron analyzer.
This analyzer uses a pre-lens system optimized for the detection of low-energy electrons, but is also capable of detecting electrons with kinetic energies up to 1.5~keV.
The main vacuum chamber contains an efficient µ-metal shielding.
All experiments were conducted with a pass energy of 200~eV, and a nominal analyzer energy resolution of 2.0~eV (first campaign, entrance-slit setting 4.0~mm) or 0.4~eV (all other measurements, entrance-slit setting 0.8~mm, straight slit).
For the liquid-water measurements, a small amount ($\approx50$~mM) of NaCl salt was added to highly demineralized water (conductivity $\approx0.2$~µS~cm$^{-1}$) to maintain electrical conductivity and mitigate potentially deleterious sample-charging effects.\cite{Kurahashi2014, Thuermer2021, Pugini}
This is common practice when measuring photoelectron spectra from liquid water.\cite{Winter2006, Pohl2019}
The liquid microjet was generated by injecting the sample solution into the interaction vacuum chamber through a 28~µm diameter glass capillary, at a typical flow rate of 0.8~mL~min$^{-1}$.
A cooling jacket extending up to approximately 70~mm upstream of the nozzle and stabilized at 10~$^\circ$C was used to reduce the temperature of the liquid sample (see Ref.~\cite{EASI} for details).
We then positioned the point of irradiation of the liquid jet at a distance of 0.5-0.8~mm from the 800~µm diameter skimmer orifice at the analyzer entrance.
The pressure in the main chamber was kept at approximately $5 \cdot 10^{-4}$~mbar using two turbo-molecular pumps (with a total pumping speed of $\approx2600$~L~s$^{-1}$ for water) and three liquid-nitrogen cold traps (with a total pumping speed of $\approx$35000~L~s$^{-1}$ for water).
The solution was delivered using a Shimadzu LC-20 AD HPCL pump, equipped with an in-line degasser (Shimadzu DGU-20A5R).
Measurements were carried out for light and heavy water, respectively.
Figure~2 shows a sketch of the experimental geometry.

\begin{figure}
\includegraphics[width=0.6\columnwidth]{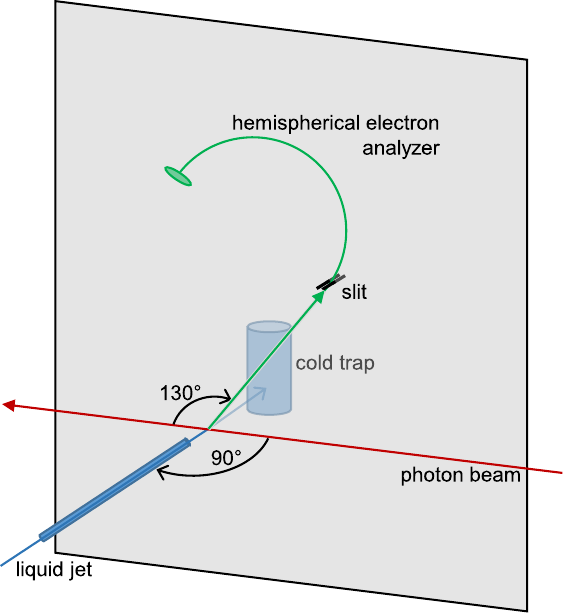}%
\caption{
Sketch of the experimental geometry. Circularly polarized synchrotron radiation (red arrow) is crossed by the liquid jet (blue arrow) in the horizontal plane under a right angle. Electrons are collected by a hemispherical analyzer arranged to have a vertical orbit plane aligned with the synchrotron radiation beam (green line: central electron trajectory). The entrance slit into the hemisphere therefore is parallel to the liquid jet. After crossing the interaction zone the liquid jet is frozen out on a cold trap.
\label{fig:setiü}}%
\end{figure}

The synchrotron-light propagation axis was orthogonal to the liquid jet, with both lying in the horizontal plane.
We used the hemispherical electron analyzer positioned at a 130$^\circ$ angle with respect to the photon-beam propagation direction (backward-detection geometry \cite{EASI}), with its lens lying in the vertical plane, and circular polarization of the photon beam.
All photoemission measurements reported here were conducted using the 1200~l~mm$^{-1}$ laminar grating of the P04 beamline.
For the first campaign, we used an exit-slit setting of 1000~µm and a resulting photon flux of 3.5$\cdot10^{13}$ photons/s; for all other measurements, depending on the exact settings of the exit slit and the grating groove depth, the photon flux amounted to 4.1-7.7$\cdot10^{12}$ photons/s.
Due to the spatial dimensions of the EASI vacuum chamber, the interaction point is located $\approx220$~mm downstream of the nominal focus position when it is mounted at the P04 beamline.
For our first experiments, this, in conjunction with the large exit-slit opening, lead to a vertical spot size of $\approx400$~µm.
In all subsequent campaigns, the matching of vertical spot size to the spatial extension of the liquid jet was improved by specific settings of the refocusing mirror unit of the beamline (pair of Kirkpatrick-Baez mirrors), thus focusing the beam to 40~µm in the vertical direction.
The horizontal spot size, extending along the liquid-jet flow direction, is 180~µm.
Approximate photon-energy resolutions under these conditions are 3.0~eV (first campaign) and 0.31~eV (all other measurements).
The photon energy was normalized by a procedure that optimizes the angle for specular reflection of the grating as a reference.
As the exact photon energy is not decisive for the conclusions of this study, no further calibration was attempted.
We chose to use a photon energy of 1400~eV to be well above the known gas-phase threshold of 1171~eV \cite{Mucke2013} and the expected similar liquid-phase threshold for water double core holes. Measuring about 200~eV above threshold represents a good compromise for photon flux and cross section.

The kinetic-energy scale of the hypersatellite Auger spectra was calibrated with respect to the maximum of the normal $KVV$ Auger spectrum of liquid water, measured in conjunction with the DCH spectra and at the same photon energy.
A value of (502.7$\pm$0.3)~eV was used for the normal Auger energy; see Supplementary Material.

\subsection{Theoretical Methods}
To mimic the Auger electron spectrum of liquid water, we calculated the corresponding Auger spectrum for the central water molecule in a water pentamer in a tetrahedral arrangement using the electronic-structure toolkit \textsf{XMOLECULE} \cite{hao_efficient_2015, inhester_xray_2016} (version 0.2-145).
With the aim of explaining the qualitative effects, we solely addressed pure $K^{-2}$ configurations.

The Auger calculations were based on the one-center approximation.\cite{siegbahn_auger_1975}
Specifically, a set of molecular orbitals was obtained for double-core-ionized configurations using the restricted Hartree-Fock method and employing the maximum-overlap method.\cite{gilbert_self-consistent_2008}

All calculations were performed using the 6-31G(d,p) basis set.\cite{ditchfield_selfconsistent_1971, hehre_selfconsistent_1972, hariharan_influence_1973}
Using the obtained orbitals, the final states of the Auger process were calculated using configuration interaction (CI) employing
all two-valence-hole--one-core (2h-1c) and selected three-valence-hole--one-core-one-particle (3h-1c1p) configurations.
The number of configurations that appear in such configuration-expansion calculations rapidly explodes with the size of the system.
For the water pentamer [\ce{(H_2O)_5}], the number of spin configurations exceeds 500,000, thus leading to a considerable computational challenge, since numerous eigenvectors have to be calculated to cover the relevant final Auger states.
To keep the calculations tractable, we had to constrain the employed configurations.
To that end, valence orbitals were localized using the Foster-Boys localization procedure.\cite{foster_canonical_1960}
Only those configurations were considered, where not more than one valence hole is located outside the central water molecule.
The idea behind this restriction is that electronic configurations, where both valence holes are not covering the central water molecule, in which the core hole is located, have vanishing Auger yield.
The described approximation is tested for a water dimer in the Supplementary Material.
The restriction of the configuration space yielded a total number of 62,084 doublet configurations.
For each considered geometry, we computed the lowest 5,000 roots to cover the relevant energy range of the Auger spectrum.

The Auger amplitudes involve the evaluation of two-electron integrals of the form

\begin{equation}
\langle a b | c k \rangle=
\int \! d^3 { r}_1
\int \! d^3 { r}_2 \,
\phi_a({\bf r}_1)
\phi_b({\bf r}_2)
\frac{1}{|\mathbf{r}_1 - \mathbf{r}_2|}
\phi_c({\bf r}_1)
\phi_k({\bf r}_2),
\label{teint}
\end{equation}

where $\phi_a ({\bf r})$ and $\phi_b ({\bf r})$ are valence orbitals, and $\phi_c ({\bf r})$ is a core orbital.
The function $\phi_k({\bf r})$ represents an energy-normalized continuum wave function with the Auger electron energy $\epsilon_k=k^2/2 = E_{\mathrm{i}} - E_{\mathrm{f}}$, where $E_i$ and $E_f$ are the initial and final bound-state energies, respectively.

The employed one-center approximation relies on the idea that the Auger effect is dominantly an inner-atomic process.
The molecular continuum wave function $\phi_k({\bf r})$ is thus approximated with the atomic continuum wave functions $\chi_{\kappa}({\bf r})$.
Further, the two-electron integrals are expanded using the linear combination of atomic orbitals employing a minimal basis set.
Only coefficients on the atom on which the core hole is located are taken into consideration.
The two-electron integral is thus approximated by

\begin{equation}
\langle ab|ck \rangle \simeq \sum_{\mu \nu \lambda \atop \text{on atom A}} C_{\mu a} C_{\nu b} C_{\lambda c} \langle \mu \nu |\lambda \kappa \rangle.
\end{equation}

Note that, for the atomic two-electron integrals $\langle \mu \nu |\lambda \kappa \rangle$, the atomic continuum energy $\epsilon_\kappa=\kappa^2/2$ for the corresponding atomic Auger transition is employed.
The basis set used for the expansion of the molecular orbitals and the atomic two-electron integrals $\langle \mu \nu |\lambda \kappa \rangle$ were computed using the atomic electronic-structure program \textsf{XATOM}.\cite{jurek_xmdyn_2016}

Even though the Auger matrix elements are evaluated based on a fully inner-atomic approximation, the present calculation still yields results where electronic holes are distributed over various atoms. This is because, in general, molecular orbitals spread over many atoms, and furthermore, configurational mixing spreads the created charge.
This leads in practice to considerable two-center contributions, where also neighboring water molecules acquire charge.

To incorporate the effects of core-hole-induced nuclear dynamics during the lifetime of the double core hole, we propagated the molecular geometry in the double-core-hole state using molecular dynamics (MD) and employing a time step of 0.1~fs for up to 7~fs (approximately 7 times the calculated DCH lifetime).
Trajectories were initiated for isolated water (\ce{H2O}), the water pentamer [\ce{(H2O)_5}], and the deuterated water pentamer [\ce{(D2O)_5}].
The MD trajectories started from initial geometries and velocities generated by sampling the neutral-ground-state vibrational Wigner distribution.
For \ce{(H_2O)_5}, we obtained the equilibrium tetrahedral geometry and the corresponding Hessian matrix from geometry optimization using GAMESS (version 2012-R2)\cite{barca_recent_2020} employing a polarization-continuum model to mimic the surrounding water environment.
In the MD calculations, we made sure that the respective core hole was always on the central water molecule.
The effects of the nuclear dynamics on the Auger spectrum were described following Ref.~\citenum{Marchenko2018} and Ref.~\citenum{Inhester2012}.
For both, pentamer [\ce{(H_2O)_5}] and isolated water (\ce{H2O}), the valence double-hole final states and the ensuing Auger spectrum were calculated for all 70 snapshots for a total number of 50 sampled trajectories, respectively.
For each time step, an instantaneous Auger spectrum was obtained by averaging over all trajectories and employing a Lorentzian line shape with a width determined by the calculated Auger decay rate $\Gamma$.
The resulting Auger spectrum was then compiled from the instantaneous spectra with the weights $\exp(-\Gamma t)$ accounting for the decay of the double-core-hole state.

\section{Results and Discussion}

\begin{figure}[h!tb]
\includegraphics{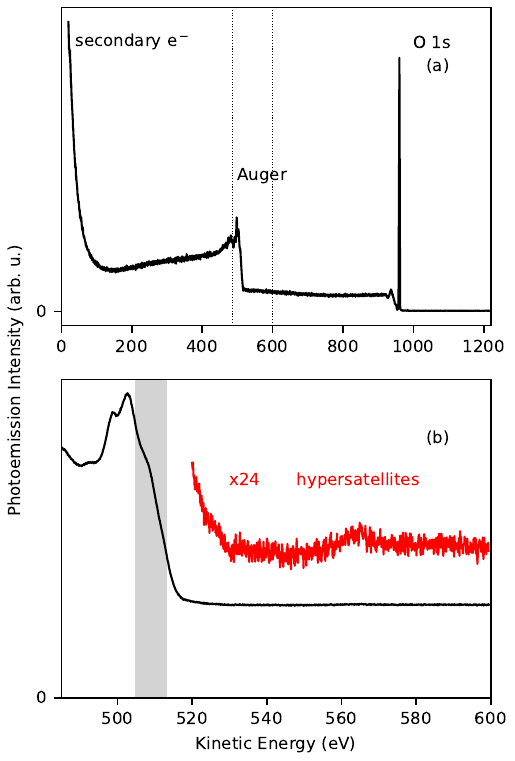}%
\caption{(a) Full experimental photoemission spectrum of liquid water, excited with photons of 1500~eV.
The kinetic-energy interval shown in the lower panel is marked by vertical dotted lines.
(b) Photoemission spectrum of liquid water showing the kinetic-energy region of Auger decay marked in panel (a), here recorded at $h\nu$ = 1400~eV.
Normal $KVV$ Auger decay can be seen between 480 and 515~eV.
The gray-shaded region from 505 to 513~eV kinetic energy is associated with proton-transferred states, for which the nuclear dynamics in the $K$-shell-ionized state plays a particular role.\cite{Thuermer2013, Slavicek2014}
Auger decay of the DCH states, 'hypersatellites', is much weaker due to the low cross section of DCH production by a single photon, and is shifted to substantially higher kinetic energies.
The upper trace in panel (b) shows a portion of the spectrum multiplied by 24 and vertically offset to improve its visibility.
We have also performed experiments below the DCH threshold and at other photon energies. These data are presented in the Supplementary Material.
\label{fig:overview}}%
\end{figure}

We start by discussing the experimental results of this study.
For reference, the inner-shell photoemission spectrum of liquid water at a photon energy above the threshold for DCH creation (1171~eV for gas-phase water)\cite{Mucke2013} is shown in Fig.~\ref{fig:overview}(a).
Although the P04 beamline is among the most brilliant synchrotron radiation sources in the soft-X-ray range, we do not expect the production of doubly ionized states by simultaneous absorption of two photons at the same or neighbouring liquid-water sites.
The spectrum therefore mostly exhibits signal from inner-shell single photoionization and the associated Auger decay.
However, the largest signal intensity is observed at the low-energy tail due to inelastically scattered electrons trailing both the O~1s main line and the $KVV$ Auger lines.
We refer to Ref.~\citenum{Malerzp1} for a general discussion of inelastically scattered electrons from liquid-water photoionization.
Auger decay of DCH states should be searched at the high-kinetic-energy side of the normal $KVV$ Auger decay, shifted by some tens of eV with respect to the decay of singly ionized states.\cite{Woods1975, Marchenko2018}
Such a spectral feature is indeed seen in Fig.~\ref{fig:overview}(b), atop of the background of inelastically scattered photoelectrons.

\begin{figure}
\includegraphics{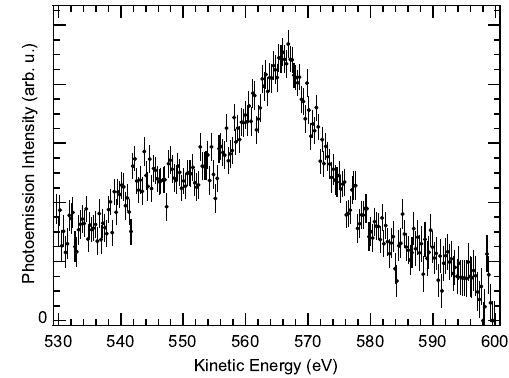}
\caption{Close-up on the photoemission intensity from liquid water on the high-kinetic-energy side of the normal Auger spectrum, recorded with a photon energy of 1400~eV.
The feature between 558 and 572~eV of kinetic energy is identified with the first step Auger decay of a DCH in liquid water.
A linear background has been subtracted.
See text for details.
\label{fig:dch}}
\end{figure}

The result obtained for much longer electron collection time and focusing on the energy region highlighted above is shown in Fig.~\ref{fig:dch}.
Here, we present the average over 111 sweeps, after subtraction of a linear background from each individual sweep.
The error bars are the standard deviation of the mean over all background-subtracted sweeps at the respective energy point.
This type of analysis clearly confirms the existence of a feature between 558 and 572~eV kinetic energy.
It was consistently observed at two different energies above the DCH creation threshold, but not at a photon energy below it (see Supplementary Material).
This feature identifies the first step in an Auger decay cascade initiated by single-photon double photoionization of liquid water.
Its intensity, relative to the normal $KVV$ Auger spectrum of singly ionized liquid water, was found as 0.0014(5), approximately 1/700 (see Supplementary Material).
This is on the same order of magnitude as found for the relative intensity of single- to double-core-hole creation in gaseous molecules.\cite{Lablanquie2016,Marchenko2018}

Our analysis of the data of Fig.~\ref{fig:dch} reveals that the main source of remaining uncertainty is not Poissonian noise from the nature of the electron-counting process, but fluctuations in the height and shape of the background.
The background exceeds the intensity of the feature under discussion by more than an order of magnitude.
Unfortunately, these fluctuations practically imposed an upper limit on the improvement in data quality we could obtain by extending the duration of the measurement (Fig.~\ref{fig:dch} corresponds to a wall-clock acquisition time of approximately 3~h).
Moreover, between different campaigns the intensity of the inelastic background and even its slope varied substantially.
More details on the raw data and their analysis are given in the Supplementary Material.

\begin{figure}
\includegraphics{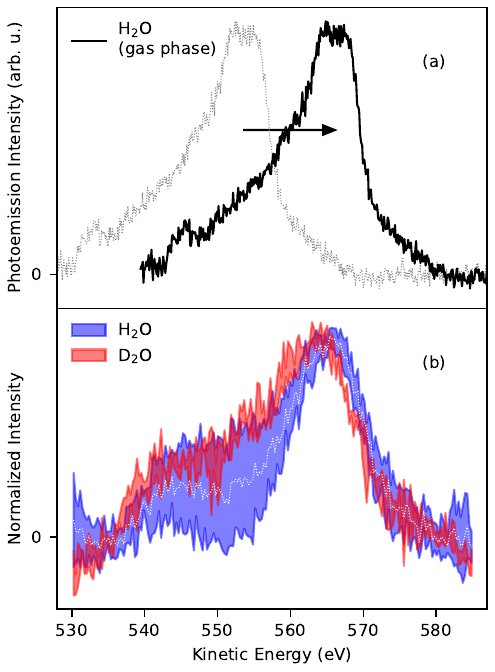}%
\caption{
Auger hypersatellite spectra of O~1s DCHs in gaseous (a) and liquid water (b).
The data for gaseous water (dotted line) were shifted by 12.5~eV (horizontal arrow) to allow for an easier comparison and were published in Ref.~\citenum{Marchenko2018}.
For the liquid, spectra were measured for normal and deuterated water.
The colored error band depicts the standard deviation of several, independently acquired data sets.
For normal water, the average is indicated by a thin dotted line.
\label{fig:expres}}%
\end{figure}

As the main result of our experiments we therefore present in Fig.~\ref{fig:expres}(b) an average of the Auger hypersatellite spectra from several data sets.
The error band now is derived from the standard deviation of the individual data sets (shown separately in the Supplementary Material), which were weighted equally to arrive at this result.
Although, in particular between 542 and 557~eV kinetic energy, strong variations in the experimental spectral shape as a function of kinetic energy can be seen, the presence of the hypersatellite signal is evident.

We can compare this signal to the one from gaseous water, shown for reference in Fig.~\ref{fig:expres}(a).
Qualitatively, we see a similar shape of the liquid-water hypersatellite Auger decay, but shifted towards higher kinetic energy by about 12.5~eV.
The shift was determined from the respective centers of gravity within the full width at half maximum of each peak.
This significant shift of the Auger spectra of liquid water relative to those of the gas phase can be rationalized qualitatively by the Born solvation model.\cite{born_1920, Barth2009}
For liquid water, the Born model is well suited to reproduce the energy shift associated with the charge state of the species under consideration (termed `M' for the sake of discussion); this can be, e.g., M$^+$ upon direct ionization and M$^{2+}$ in the case of a normal Auger process.
Here, we extend it to M$^{3+}$ in the present case of DCH formation.
Accordingly, the creation of an additional charge in a molecule embedded in a solution induces a polarization, and the resulting energy shift for a given charge state relative to the situation in vacuum amounts to
\begin{equation}
    \Delta E = \frac{q^2}{2R} \left(1-\frac{1}{\epsilon}\right),
\end{equation}
where atomic units are employed, $R$ is the assumed radius of the ionized molecule with charge $q$, and $\epsilon$ is the dielectric constant of the solution.
Following Ref.~\citenum{lundholm_core_1986} and since we have to consider mainly the electronic polarization, we employ the high-frequency dielectric constant $\epsilon=1.8$ and an ion radius estimated from the crystal radius in ice of $R=1.38$~{\AA}.
Those parameters have been found to qualitatively describe $K$-shell binding energy shifts from vapor to liquid.\cite{lundholm_core_1986}
Using the same model, we obtain a relative shift of the doubly charged double-core-hole state to the triply charged state after the hypersatellite Auger decay of about 11.6~eV, in fair agreement to the observed shift of about 12.5~eV between the liquid and vapor $K^{-2}$ Auger spectra in Fig.~\ref{fig:expres}.
Furthermore, when applying the same simple model to the single-core-hole case, we obtain a shift of 6.95~eV between gas-phase and liquid-phase Auger electrons.
This agrees reasonably well with the experimentally observed value of $\approx$5~eV.\cite{Winter2007a, Thuermer2013}

Comparing again the results for gaseous water in Fig.~\ref{fig:expres}(a) to those for liquid water in Fig.~\ref{fig:expres}(b), we observe that particularly large deviations between the various liquid-water data sets (seen as width of the error band) occur in kinetic-energy regions where contributions of the original, unshifted gas-phase signal would be expected.
To assure that the gas-phase signal makes a rather minor contribution, we have recorded reference spectra, before or after measuring the hypersatellite Auger electron spectra. This reveals a gas-phase signal contribution of less than 10\% (see Supplementary Material), far smaller than required to explain the deviations between different data sets contributing to Fig.~\ref{fig:expres}(b).
Whether the gas-phase contributions drifted towards a higher-intensity fraction during the lengthy hypersatellite data acquisition remains unsolved.

We now return to the original question of the dynamics of core-hole states in water on ultrashort time scales.
For that we have recorded the hypersatellite Auger spectra for normal and deuterated water [Fig.~\ref{fig:expres}(b)].
A difference between the spectra can be clearly seen, most significantly in the position of the high-kinetic-energy flank of the hypersatellite Auger peak.
Qualitatively, this is reminiscent of the differences between the normal $KVV$ Auger spectra of liquid light and heavy water.\cite{Thuermer2013}

\begin{figure}
\includegraphics[width=\columnwidth]{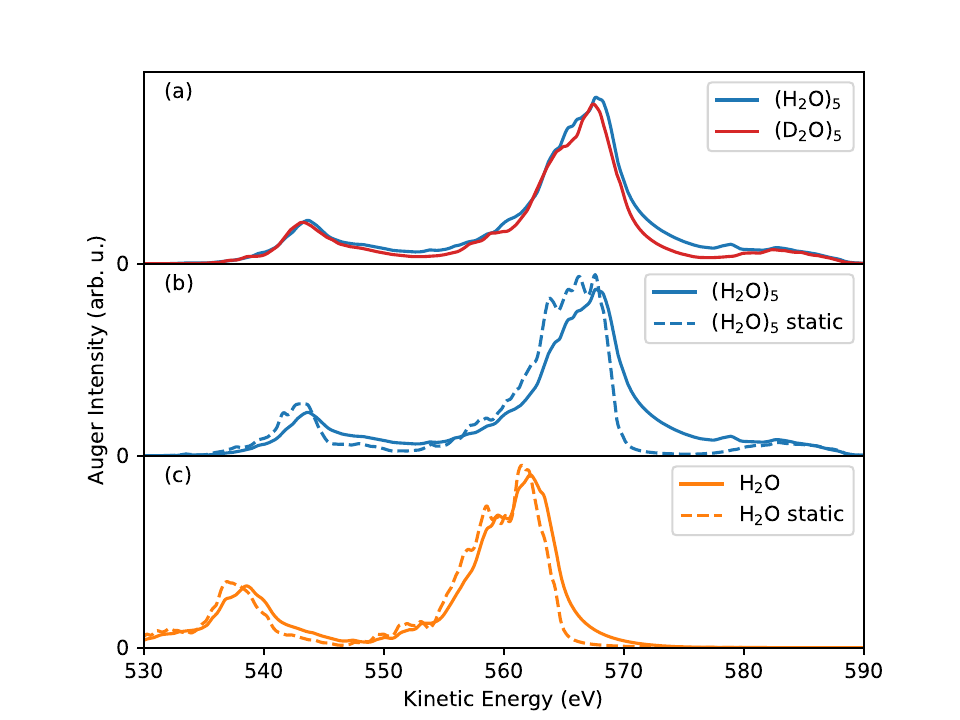}
\caption{Calculated Auger electron spectrum.
(a) Comparison for the deuterated [\ce{(D2O)_5}] and non-deuterated [\ce{(H2O)_5}] water pentamer in a tetrahedral arrangement.
(b) Auger spectrum for the water pentamer [\ce{(H2O)_5}].
(c) Auger spectrum for isolated water \ce{(H2O)}.
The dashed lines in panels (b) and (c) show instantaneous spectra at $t=0$, i.e., without taking into account the effect of core-hole-lifetime nuclear dynamics.
\label{fig:theoryCompareAuger}}
\end{figure}

Our results can be interpreted with assistance by the theoretically calculated $K^{-2}$ Auger spectra.
In order to capture the nuclear dynamics of the process, we have modeled the liquid sample by calculating Auger spectra for the central water molecule inside a water pentamer with tetrahedral arrangement, \ce{(H_2O)_5}, and its deuterated variant \ce{(D_2O)_5}.
Figure~\ref{fig:theoryCompareAuger}(a) shows a comparison of the calculated Auger spectra obtained for \ce{(H_2O)_5} and \ce{(D_2O)_5}.
The spectra exhibit two dominant peaks at 565~eV and at 545~eV that can be associated with Auger transitions leaving two vacancies in the outer-valence levels (\ce{3a1}, \ce{1b2}, and \ce{1b1}) or one vacancy in the outer- and one in the inner-valence levels (\ce{2a1}).
Despite the fact that only a single hypersatellite initial state ($K^{-2}$) was taken into account, whereas the experimental data may involve also other excited states based on a DCH (i.e., $K^{-2}V$, $K^{-2}L^{-1}$, or $K^{-2}L^{-1}V$),\cite{Marchenko2018} the calculated Auger spectrum shows qualitatively similar features as the experimental data.
The comparison of the two calculated spectra for the deuterated and the non-deuterated pentamer also shows similar trends.
As in the case of the experimental spectra, for \ce{(H_2O)_5}, the calculated high-energy flank is shifted to somewhat higher energies compared to the spectrum for \ce{(D_2O)_5}.
In addition, for the calculated spectra, the extended tail at about 580~eV tends to be more pronounced for \ce{(H_2O)_5} than for \ce{(D_2O)_5}.
This difference in the Auger spectrum can be attributed to the ultrafast core-hole-induced proton dynamics, which is less pronounced for \ce{(D_2O)_5} due to the larger mass.
The effect of the proton dynamics is highlighted by comparison
to the calculated Auger spectrum for \ce{(H_2O)_5} employing a static geometry.
As can be seen, the proton dynamic results in the pronounced tail and a slight shift of the high-energy flank to higher energies for the main Auger peak.
The same trends appear for a water molecule in the gas phase, as shown in Fig.~\ref{fig:theoryCompareAuger}(c).\cite{Inhester2012, Marchenko2018, ismail2024ultrafast}

Effects specific to the liquid environment can be understood by comparing the Auger spectrum for \ce{(H_2O)_5} in Fig.~\ref{fig:theoryCompareAuger}(b) with the one calculated for isolated water that is shown in Fig.~\ref{fig:theoryCompareAuger}(c).
The most striking difference is the significant shift of the entire spectrum to higher energies by about 5~eV.
As described above, this shift can be attributed to the fact that the electronic structure of the surrounding water molecules contributes to the final double-valence-hole states, effectively lowering the pentamer's energy through polarization.
The shift revealed from the experiment in Fig.~\ref{fig:expres} with about 12.5~eV is considerably larger than the theoretical one, which we mainly attribute to the fact that the calculation is for a water pentamer and not the full liquid environment.
In addition, this discrepancy might also arise from the relatively small basis set that had to be chosen to make the computations feasible. Remember that the experimental spectra are in good agreement with the energy shift obtained from the simple Born model.
We next comment on an additional spectral feature in Fig.~\ref{fig:theoryCompareAuger}, occurring near 580~eV kinetic energy for (H$_2$O)$_5$ and (D$_2$O)$_5$.
This identifies final double-valence-hole states with charges delocalized over neighboring water molecules and thereby results in a particularly high energy.
These channels can be best characterized as core-level intermolecular Coulombic decay,\cite{Jahnke2020} where electrons on the neighboring molecules participate in the radiationless decay.
The feature connected with this decay process amounts to 6\% of the overall decay.
We expect that this pentamer feature becomes more smeared out in liquid water due to the stronger fluctuations in the geometrical arrangements at room temperature and effects from the second solvation layers.
It is thus not surprising that this feature would be difficult to detect in the experimental data in Fig.~\ref{fig:expres}.

Apart from the overall shift caused by polarization and the additional peak at 580~eV linked to decays directly involving the neighboring water molecules, the calculations show further effects of the local chemical environment.
A comparison with the respective static spectra in Figs.~\ref{fig:theoryCompareAuger}(b) and \ref{fig:theoryCompareAuger}(c) exhibits that the effects of the nuclear dynamics in \ce{(H2O)_5} are significantly stronger.
Specifically, the marked tail on the high-energy side of the dominant Auger line located at 565~eV for (H$_2$O)$_5$ and at 560~eV for H$_2$O is more pronounced for (H$_2$O)$_5$ compared to H$_2$O.

\begin{figure}
\includegraphics[width=\columnwidth]{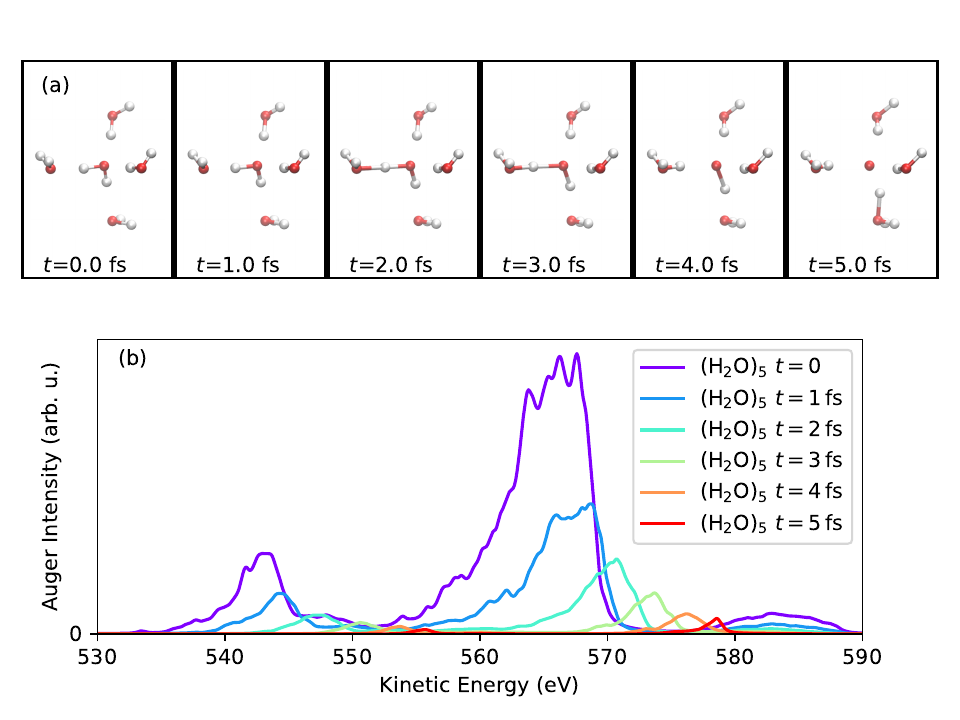}
\caption{
(a) Snapshots of the molecule geometry for a typical \ce{(H2O)_5} trajectory.
(b) Calculated instantaneous Auger spectra at selected times after core ionization.
\label{fig:theoryInstantaneous}}
\end{figure}

\begin{figure}
\includegraphics[width=\columnwidth]{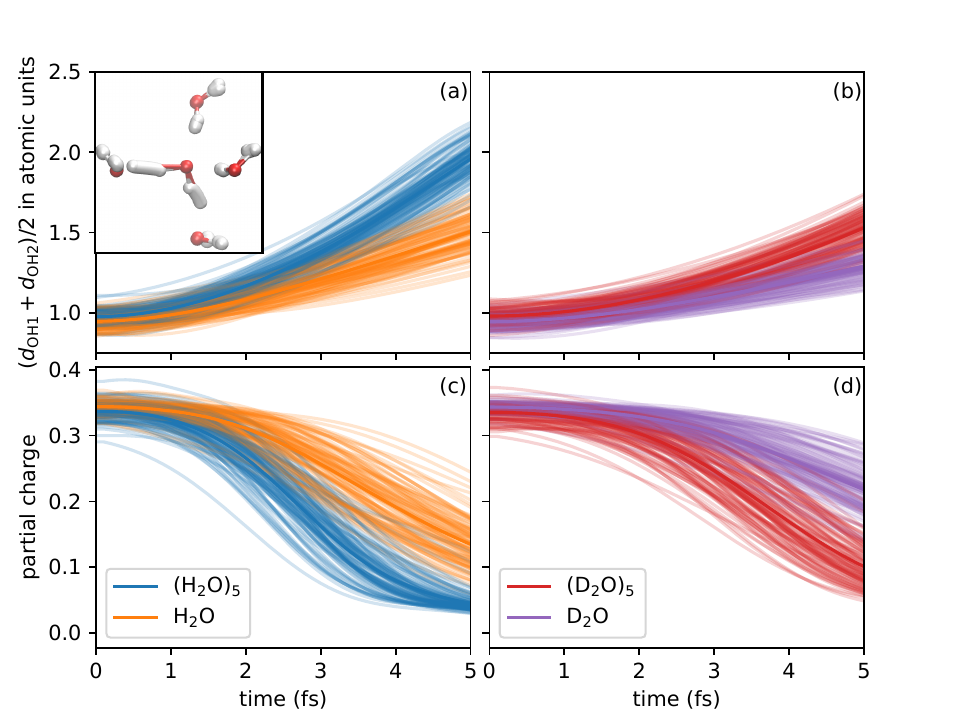}
\caption{Top panels: Symmetrized oxygen-hydrogen distance as function of time after double core ionization.
(a) Comparison of water pentamer in a tetrahedral arrangement [\ce{(H2O)_5}] and isolated water (\ce{H2O}).
The inset illustrates an example trajectory for the pentamer [see Fig.~\ref{fig:theoryInstantaneous}(a) for individual snapshots].
(b) Comparison of deuterated water pentamer [\ce{(D2O)_5}] and isolated deuterated water.
Bottom panels: Mulliken charge on the central oxygen as a function of time. Comparison for (H$_2$O)$_5$ and H$_2$O (c) as well as (D$_2$O)$_5$ and D$_2$O (d).
\label{fig:theoryCompareDistances}}
\end{figure}

The effect of the nuclear dynamics on the spectrum is further discussed in Fig.~\ref{fig:theoryInstantaneous}, where instantaneous Auger spectra at selected times are shown.
The individual instantaneous spectra are weighted with their relative contribution according to the exponential decay in the core-hole lifetime of $\approx$1.5~fs in the calculations.
In the double-core-hole state, the water molecule is highly dissociative, and both hydrogen atoms dissociate as protons.\cite{inhester_xray_2016}
The dynamics is sketched for a typical trajectory in Fig.~\ref{fig:theoryInstantaneous}(a).
The repulsion of the two protons reduces the charge around the oxygen atom [see Figs.~\ref{fig:theoryCompareDistances}(c) and \ref{fig:theoryCompareDistances}(d) for details], ultimately leading to faster Auger electrons.
As can be seen in Fig.~\ref{fig:theoryInstantaneous}(b), 
the spectral shifts result in marked tails on the high-energy side of the Auger peaks.

For the double-core-hole state, asymmetric stretching and bond-angle opening play a minor role during the lifetime of the Auger decay, instead the dominant dynamics for the water pentamer as well as the isolated water molecule occurs along the symmetric stretching leading to the transfer of two protons to the neighboring molecules.
This is different for the single-core-ionized water molecule, where, in the liquid phase, dynamics involves the transfer of a \textit{single} proton \cite{Thuermer2013, Slavicek2014} and, in the gas phase, the molecule remains stable.\cite{Inhester2012}
Notably, as shown by Marchenko et al. \cite{Marchenko2018}, we expect that for other double-core-hole satellite configurations, asymmetric stretching and bond-angle opening may also play a substantial role.\cite{Marchenko2018}
Moreover, we point out that proton dynamics for other double-core-hole satellites can be significantly faster compared to the pure DCH state, as it was reported for the gas phase.\cite{Marchenko2018}
Figure~\ref{fig:theoryCompareDistances} displays the geometrical variations of the trajectories used to compute the Auger spectra.
It shows the symmetrized bond distance of the central water molecule of \ce{(H_2O)_5} as a function of time after core ionization.
To make the difference in the dynamics more apparent, we plot the data for up to 5~fs, but one has to keep in mind that the decay lifetime of the double core hole is $1/\Gamma \simeq \unit[1.5]{fs}$.
Figure~\ref{fig:theoryCompareDistances}(a) reveals that the fragmentation of the water molecule in the double-core-hole state is somewhat faster in the tetrahedral water structure of \ce{(H_2O)_5} than in \ce{H_2O}.
Even though the effect is rather small at early times, it points towards the fact that the liquid environment tends to accelerate the proton dynamics compared to the case of an isolated water molecule.
This result can also be linked to the stronger effect of nuclear dynamics in \ce{(H_2O)_5} vs. \ce{(H_2O)} seen in Fig.~\ref{fig:theoryCompareAuger}(b).
This result is plausible, considering that the presence of hydrogen bonds in \ce{(H_2O)_5} promotes the proton dissociation relative to the situation in \ce{H_2O}.

Figure~\ref{fig:theoryCompareDistances}(b) shows the dynamics for \ce{(D_2O)_5} and for \ce{(D_2O)}.
As one might expect, the dissociation for \ce{(D_2O)_5} is somewhat slower than for \ce{(H_2O)_5}.
When comparing the respective deuterated and non-deuterated species, one can see that deuteration leads to a similar deceleration of the proton dynamics as the transition from \ce{(H_2O)_5} to \ce{H_2O}.
Our results show that the ultrafast core-hole-induced dynamics is promoted when the double-core-ionized water molecule is in a tetrahedral hydrogen-bond environment.
We speculate that this trend also holds for a liquid-water environment.
Figures~\ref{fig:theoryCompareDistances}(c) and \ref{fig:theoryCompareDistances}(d) show the partial charge (Mulliken charge) of the central oxygen atom and how it evolves as a function of time for (H$_2$O)$_5$, H$_2$O [Fig.~\ref{fig:theoryCompareDistances}(c)], (D$_2$O)$_5$, and D$_2$O [Fig.~\ref{fig:theoryCompareDistances}(d)].
The partial charge of the central oxygen atom is initially about 0.35 due to the strong screening effects that involve electronic rearrangements from neighboring atoms towards the double-core-ionized oxygen atom.
As the protons move away from the central oxygen atom, the partial charge assigned to the central oxygen atom decreases, illustrating the above-mentioned reduction of charge via proton repulsion.

We have tried to extract further experimental information on DCHs in liquid water by attempting to observe primary electrons corresponding to resonant $K^{-2}V$ satellite states at binding energies around the DCH threshold; no such signature could be identified, however (see Supplementary Material).

%

\section{Summary and Outlook}
We experimentally characterized the Auger electron hypersatellite spectrum from the first step in the radiationless decay cascade of single-site DCHs in liquid water, produced by single-photon double photoionization.
Compared to its gas-phase counterpart, the kinetic energy of the feature is blue-shifted by 12.5~eV, largely due to polarization of the surrounding water molecules in the final state.
Despite the lifetime of an oxygen DCH being around 1.5~fs, nuclear dynamics within this time window is evidenced by a noticeable isotope effect when experiments are carried out with deuterated water.
Calculations of the Auger spectrum of a DCH in a water pentamer allowed to identify a symmetric stretch of the molecule as the main driver of nuclear dynamics and shed light on the temporal evolution of the Auger spectrum.

As an outlook we would like to present some thoughts about future experiments on DCHs.
Studies of the nuclear dynamics during normal Auger decay and the spectral features associated with it (see, e.g., Refs.~\citenum{Fuchs2008, Tokushima2008, Odelius2009} and the review Ref.~\citenum{Fransson2016}) had considerable impact on the discussion about the structure of water.
Extending these studies to DCHs will extend the range of parameters under which this dynamics can be probed.

One important aspect for the quality of DCH-related electron spectra we have identified is the ratio of the signal of interest to background from inelastic scattering of O~1s photoelectrons.
This might be more favourable at higher photon energies, for two reasons:
(1) The background will be stretched out to a wider kinetic-energy range.
(2) The relative cross section of double- to single-core-hole creation depends on the excess energy.
This is brought about by an interplay of two mechanisms for single-photon double photoionization that were discussed, e.g., in Ref.~\citenum{Rost2002}.
If we assume this discussion to be universal, the maximum of the relative DCH production should be at an excess energy of roughly 1.4 times the binding energy of the DCH state, i.e., for our system at a photon energy of around 1700~eV.
Experimentally, for Ne, isoelectronic to water, an increase of the double/single core-hole-production ratio was found up to photon energies of 5~keV.\cite{Southworth2023}
We believe that these factors can lead to hypersatellite Auger spectra of better quality at higher photon energies, even though photon flux and absolute cross sections might be lower than in the present study.
Moreover, we suggest liquid-water DCH experiments at XFELs, where the ratio between $K^{-2}$ and $K^{-1}$ should be larger and we would be less sensitive to an unknown and varying background.

As a further outlook, we want to point out that nowadays measuring time-resolved Auger spectra is in reach.\cite{Haynes2021}
So, as discussed above, our calculations reported here are interesting not only regarding the high-energy tail, but also the energy shift (see Fig.~\ref{fig:theoryInstantaneous}).
We envision time-resolved Auger spectra using intense X-ray free-electron lasers and, e.g., self-referenced attosecond streaking \cite{Haynes2021} of more complex systems than noble gases in the future.
So far, e.g., the double-core-hole generation in O$_2$ molecules and the corresponding molecular-frame photoelectron angular distributions \cite{Kastirke2020} and an alternative pathway to double-core-hole states in gas-phase water \cite{Ismail2023} have been studied at European XFEL.

Finally, we would like to mention that double-core-hole production has also been studied for anions, partly giving rise to significantly larger cross sections than in neutrals.\cite{Schippers2020}
For example, the maximum cross section of (1s+1s) ionization in C$^-$ has been found as 3~kb,\cite{Perry2020} compared to 0.07~kb when our experimental double/single core-ionization ratio is combined with a calculated value for the O~1s photoionization cross section.\cite{Yeh}
No explanation of this qualitative difference is known to the authors.
Experimentally, the study of systems that in solution are present in anionic form is well possible, since in aqueous phase anions can be readily stabilized.
Therefore, the search for a potential enhancement of the DCH cross section with charge state seems highly interesting.


%
%

%

\section*{Supplementary Material}
In the Supplementary Material, we first describe details of the acquisition of the experimental data and their analysis. We then comment on various aspects of our results that may be of interest to a specialized reader.

\begin{acknowledgments}
We thank Claudia Kolbeck for her contributions to data acquisition.
We acknowledge DESY (Hamburg, Germany), a member of the Helmholtz Association HGF, for the provision of experimental facilities.
Parts of this research were carried out at PETRA~III and we would like to thank Moritz Hoesch and his team for assistance in using beamline P04.
Beamtime was allocated for proposals I-20180132, H-20010092, II-20180012, and I-20200356.
F.T.\ acknowledges funding by the Deutsche Forschungsgemeinschaft (DFG, German Research Foundation) - Project 509471550, Emmy Noether Programme.
F.T.\ and B.W.\ acknowledge support by the MaxWater initiative of the Max-Planck-Gesellschaft.
L.I.\ acknowledges support by the Cluster of Excellence `CUI: Advanced Imaging of Matter' of the Deutsche Forschungsgemeinschaft (DFG) - EXC 2056 - project ID 390715994 and the scientific exchange and support of the Centre for Molecular Water Science (CMWS).
S.T.\ acknowledges support from the JSPS KAKENHI Grant No. JP20K15229.
T.M.\ acknowledges funding from the European Union’s Horizon 2020 research and innovation program under the Marie Skłodowska-Curie grant agreement No.\ 860553.
B.W.\ and U.H.\ acknowledge funding from the European Research Council (ERC) under the European Union's Horizon 2020 research and innovation program under grant agreement No.\ 883759.
\end{acknowledgments}

\section*{Author declarations}
\subsection*{Conflict of Interest}
The authors have no conflicts to disclose.
\subsection*{Author Contributions}
Conceptualization: R.P., T.M., M.N.P., and M.S.;
Data curation: U.H.;
Formal Analysis: U.H.;
Investigation: F.T., S.M., S.T., B.W., and U.H.; 
Methodology: F.T., L.I., and U.H.;
Visualization: L.I. and U.H.;
Project administration: B.W. and U.H.;
Supervision: B.W.;
Funding acquisition: B.W. and U.H.;
Writing – original draft: F.T., L.I., and U.H.;
Writing – review \& editing: All authors.

\section*{Data Availability}
The data that support the findings of this study are openly available in the Zenodo data repository at https://doi.org/10.5281/zenodo.10523680, Ref.~\citenum{dataset}.

\bibliography{main}

\end{document}



\title{Radiationless Decay Spectrum of O~$1s$ Double Core Holes in Liquid Water --
Supplementary Material} 


\author{Florian~Trinter}
\email{trinter@fhi-berlin.mpg.de}
\affiliation{Fritz-Haber-Institut der Max-Planck-Gesellschaft, Faradayweg~4-6, 14195 Berlin, Germany}
\affiliation{Institut für Kernphysik, Goethe-Universität Frankfurt, Max-von-Laue-Str. 1, 60438 Frankfurt am Main, Germany}

\author{Ludger~Inhester}
\email{ludger.inhester@cfel.de}
\affiliation{Center for Free-Electron Laser Science CFEL, Deutsches Elektronen-Synchrotron DESY, Notkestr. 85, 22607 Hamburg, Germany}

\author{Ralph~Püttner}
\affiliation{Fachbereich Physik, Freie Universität Berlin, Arnimallee~14, 14195 Berlin, Germany}

\author{Sebastian~Malerz}
\affiliation{Fritz-Haber-Institut der Max-Planck-Gesellschaft, Faradayweg~4-6, 14195 Berlin, Germany}

\author{Stephan~Thürmer}
\affiliation{Department of Chemistry, Graduate School of Science, Kyoto University, Kitashirakawa-Oiwakecho, Sakyo-Ku, 606-8502 Kyoto, Japan}

\author{Tatiana~Marchenko}
\affiliation{Sorbonne Université, CNRS, Laboratoire de Chimie Physique -- Matière et Rayonnement, LCPMR, 75005 Paris, France}

\author{Maria~Novella~Piancastelli}
\affiliation{Sorbonne Université, CNRS, Laboratoire de Chimie Physique -- Matière et Rayonnement, LCPMR, 75005 Paris, France}

\author{Marc~Simon}
\affiliation{Sorbonne Université, CNRS, Laboratoire de Chimie Physique -- Matière et Rayonnement, LCPMR, 75005 Paris, France}

\author{Bernd~Winter}
\affiliation{Fritz-Haber-Institut der Max-Planck-Gesellschaft, Faradayweg~4-6, 14195 Berlin, Germany}

\author{Uwe~Hergenhahn}
\email{hergenhahn@fhi-berlin.mpg.de}
\affiliation{Fritz-Haber-Institut der Max-Planck-Gesellschaft, Faradayweg~4-6, 14195 Berlin, Germany}

\date{\today}

\begin{abstract}
\centering 
In the Supplementary Material, we firstly delineate details of the acquisition of the experimental data and their analysis.
Subsequently, we comment on various aspects of our results that might be of interest for a specialized reader.
\end{abstract}

\maketitle 
\renewcommand{\figurename}{Supplementary Fig.}
\section{Experimental Methods}
\subsection{Data acquisition}
Spectra were recorded with our general-purpose liquid-jet photoemission setup EASI, see Ref.~\cite{EASI}, at the P04 beamline \cite{P04} of the synchrotron radiation source PETRA~III (DESY, Hamburg, Germany).
For normal water (H$_2$O), five independent data sets were recorded within a time span of almost three years; for deuterated water (D$_2$O), two independent data sets were recorded in different beamtimes.

\begin{figure}[htb!]
\includegraphics{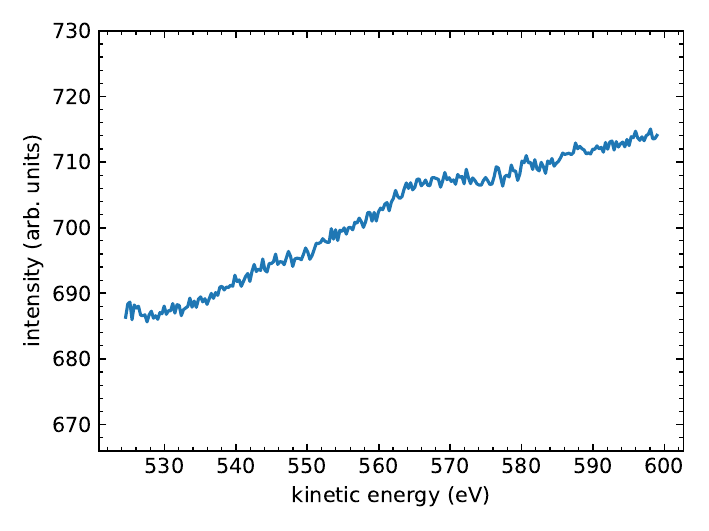}%
\caption{Photoemission spectrum of water in the energy region of O~$1s^{-2}$ radiationless decay, recorded at 1400~eV photon energy. Average over twenty sweeps.
\label{sfig:avgtrace}}%
\end{figure}

Due to the weak signal of the double-core-hole decays compared to other phenomena driven by single photoionization, data acquisition over extended periods of time was necessary.
Typically, twenty to more than hundred sweeps over the relevant energy region (see Fig.~4 of the main article) were recorded and averaged over.
A typical spectral trace, averaged over twenty sweeps, is shown in Supplementary Fig.~\ref{sfig:avgtrace}.
In total, the data shown in the figure correspond to approximately 1200~s of acquisition.

\begin{figure}[htb!]
\includegraphics{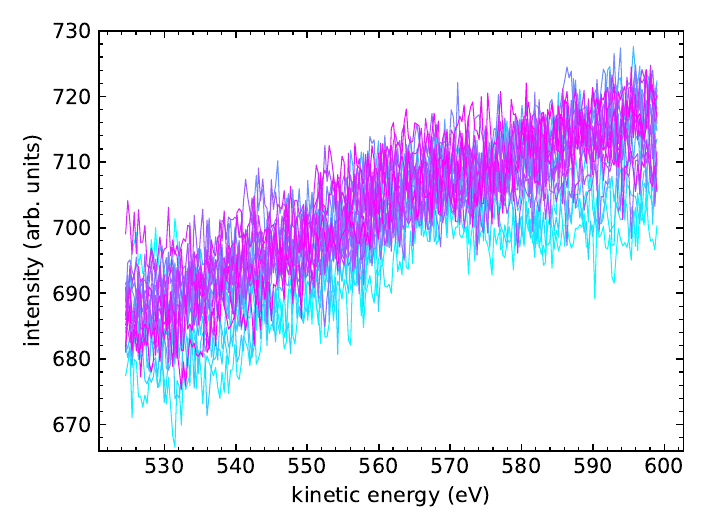}%
\caption{Photoemission spectrum of water recorded at 1400~eV photon energy, twenty individual sweeps recorded subsequently (color scale: cyan to magenta in temporal order).
\label{sfig:traces}}%
\end{figure}

Electron intensity at the end of the analyzer is measured with a microchannel plate plus CCD camera combination.
To increase the dynamic range, the gray-scale levels of the CCD camera are converted into intensity levels (`ADC mode').
By that, the statistics of the data does not correspond to a counting process, and the noise level is not adequately described by Poissonian statistics.
Nevertheless, more important than shot noise from the counting process are intensity fluctuations that are present in the spectra from a combination of factors not separated in detail.
This can well be seen when the twenty sweeps making up the average in Supplementary Fig.~\ref{sfig:avgtrace} are considered individually (Supplementary Fig.~\ref{sfig:traces}).
Plausibly, fluctuations of the jet position on a µm-scale contribute to this level of sweep-to-sweep stability.
We note that the signal in the energy region shown here is mostly comprised of a background of inelastically scattered electrons, as described in the main text.

\begin{figure}[htb!]
\includegraphics{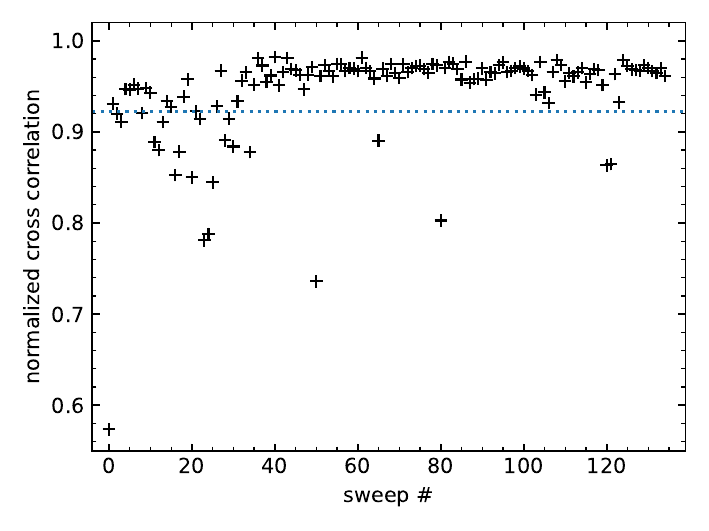}%
\caption{Normalized cross correlation between a single sweep and the average of 135 sweeps over the energy range shown in Supplementary Figs.~\protect\ref{sfig:avgtrace} and \protect\ref{sfig:traces}, as a function of sweep number.
Sweeps were discarded below a cutoff of 0.922 in this example (dotted blue line).
\label{sfig:ccorrelation}}%
\end{figure}

\subsection{Sweep-to-sweep stability and data selection}
In Supplementary Fig.~\ref{sfig:traces}, most sweeps seem to have a common shape, but some outliers can be identified.
In order to exclude outlier sweeps from the further analysis, for each set of data we determined the normalized cross correlation for every individual sweep with the average over all sweeps within the data set (see Supplementary Fig.~\ref{sfig:ccorrelation}), defined here as:
\begin{equation}
 \mathrm{ncor}\bigl(a,b^{(j)}\bigr) = \biggl({\sum_i a_i\, b_i^{(j)}}\biggr)\,{N}^{-1},
\label{eq:ccross}
\end{equation}

with the following notation: $N$ the number of data points, $K$ the number of sweeps; 
$t_i^{(j)}$ (a real, positive number) the {\it i}th energy data point of the {\it j}th trace, with $i = 1\dots N, j = 1\dots K$; $t^{(j)}$ all $N$ data points of the $j$th trace; $\langle t^{(j)}\rangle$ the mean over the $N$ data points of the $j$th trace;
$\tilde t_i = \langle t_i^{(j)}\rangle_{j = 1\dots K}$ the mean over the $K$ traces at the $i$th data point; $\tilde t$ all $N$ data points of the sweep-averaged trace;
$\langle\tilde t\rangle$ the mean over the $N$ data points of the sweep-averaged trace. We then use %

\begin{equation}
a_i = \frac{\tilde t_i - \langle \tilde t\rangle}{\sigma (\tilde t)},\qquad
b_i^{(j)} = \frac{t_i^{(j)} - \langle t^{(j)}\rangle}{\sigma (t^{(j)})},
\end{equation}
with
\begin{equation}
\sigma(t) = \biggl( \sum_i \bigl|t_i - \langle t\rangle\bigr|^2\,N^{-1}\biggr)^{1/2}
\end{equation}
the standard deviation of the trace $t$.
This metric will result in a normalized cross correlation of unity for traces that agree in shape, even if one of them suffers from an overall loss of intensity. 
On the other hand, reduced values of $\mathrm{ncor}(a,b)$ are produced for traces $a,b$ of different shape even if their intensity is similar.
Overall loss of intensity was observed in some of the campaigns analyzed here when spectra were recorded over hours without re-adjusting the liquid-jet position.
This problem was greatly reduced in later runs by changing to a more rigid 3D manipulator holding the liquid jet.

\begin{figure}[htb!]
\includegraphics{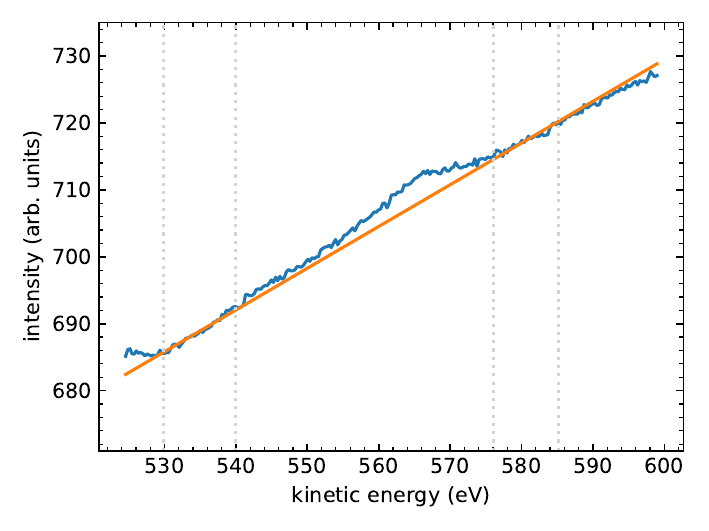}%
\caption{Averaged water photoemission spectrum (blue), produced from all sweeps above the cross-correlation cutoff shown in Supplementary Fig.~\protect\ref{sfig:ccorrelation}. 
A linear background function (orange) has been fitted to the portions of the spectrum between each pair of dotted lines.
\label{sfig:bg}}%
\end{figure}

\subsection{Background subtraction}
From the average of all traces matching the cross-correlation criterion, a linear background was subtracted (Supplementary Fig.~\ref{sfig:bg}).
We note at this point that the slope of the bulk of secondary electrons as a function of energy varied between the different data-acquisition campaigns. 
This led to a varying slope of the linear function that was subtracted from the data; the procedure described in the caption of Supplementary Fig.~\ref{sfig:bg} was applied in all cases.
The true background below the feature will most probably exhibit some curvature. We nevertheless stayed with the simple linear model because we felt that fitting more sophisticated functions to the fluctuating signal in the background regions shown in the figure might produce non-physical variations between the different data sets.

\begin{figure}[htb!]
\includegraphics{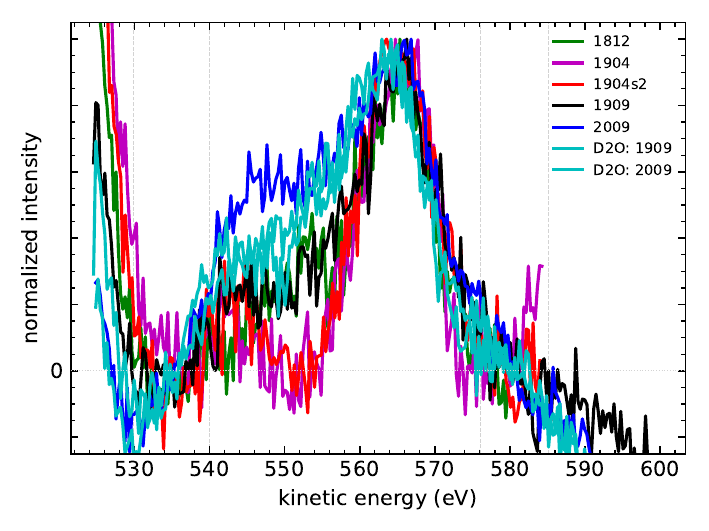}%
\caption{Background-subtracted electron intensity traces for water photoemission, all data sets acquired for this study.
All traces were normalized to equal maximum intensity of the double-core-hole decay feature (hypersatellites).
Data sets are labelled by the time of recording in {\it yymm} format.
The regions used for background subtraction are marked by dotted vertical lines, see Supplementary Fig.~\protect\ref{sfig:bg}.
\label{sfig:alldata}}%
\end{figure}

\subsection{Comparison of the data from different campaigns}
Background-subtracted traces from all data sets are shown in Supplementary Fig.~\ref{sfig:alldata}.
As there was no way to calibrate the intensity between the different acquisition campaigns, all traces are normalized to equal maximum height. 
From the figure it is clear that (1) intensity can reproducibly be seen in the energy region expected for radiationless decay of double core holes, when results for gas-phase water \cite{Marchenko2018} are shifted upwards on the kinetic-energy scale, which is expected due to final-state polarization, (2) besides the main double-core-hole decay peak substantial variations of the signal height can be observed, which outweigh the statistical fluctuations of the background-subtracted data.
The precise origin of this intensity is unknown at this moment, but since in this energy region the signal by scattered electrons is far stronger than the decay signal of the core holes, and might easily depend on apparative factors to some degree, it seems reasonable to attribute the observed instability to the background signal.
As these variations due to background instabilities are the biggest source of data error, we consider it most adequate to treat all data sets equally, independent of the fact that the acquisition times of the individual sets varied between two and eight hours.
The same background regions were used for all data sets, although the acquired energy range varied slightly.
Because of that, the linear background applied as shown in Supplementary Fig.~\ref{sfig:bg} for some data sets is slightly overestimated and leads to negative intensities in some energy regions that are not of interest for the discussion.
\begin{figure}[htb!]
\includegraphics{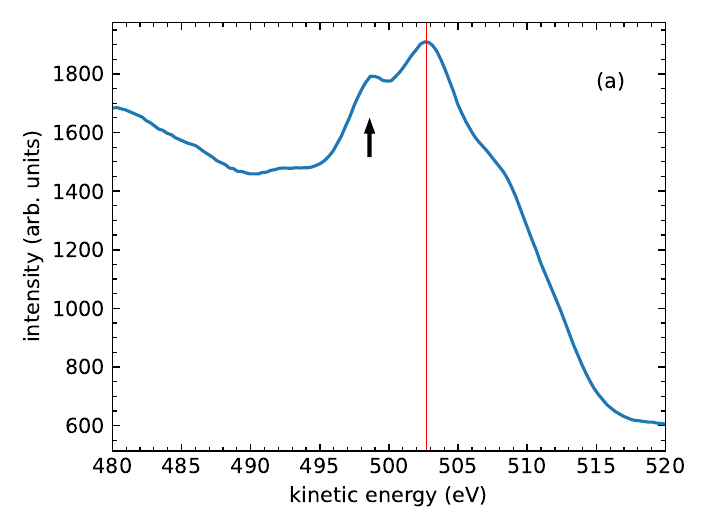}\\
\includegraphics{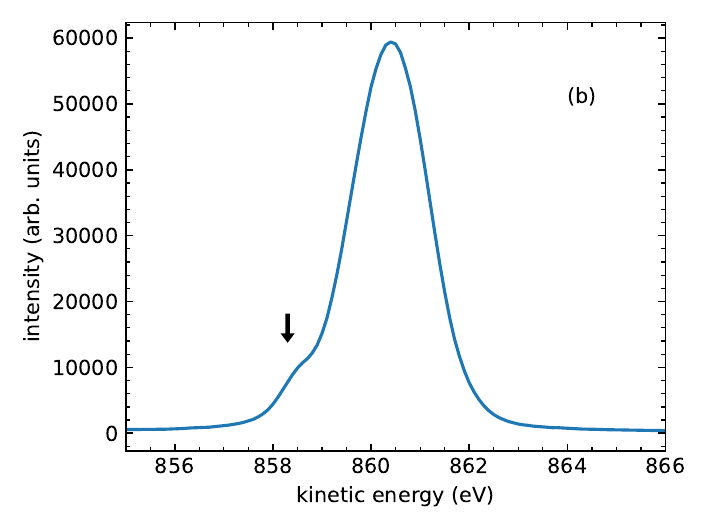}%
\caption{
Normal KVV Auger spectrum [panel (a)] and O~$1s$ photoelectron spectrum [panel (b)] recorded under the same conditions as the Auger hypersatellite spectra.
The positions of features due to residual gas-phase water surrounding the jet are marked by arrows.
A thin vertical line (red) in panel (a) marks the position of the maximum at 502.7(3)~eV kinetic energy.
\label{sfig:lqgas}}%
\end{figure}
\subsection{Kinetic-energy calibration}
For all Auger hypersatellite spectra, the kinetic-energy scale of the analyzer was calibrated to KVV Auger spectra of normal liquid water \cite{Thuermer2013} measured immediately before or after the hypersatellite spectra at the same photon energy of 1400~eV.
The position of the maximum of the spectrum was used as reference (see Supplementary Fig.~\ref{sfig:lqgas}).
As a prerequisite, we determined its value within this work as described below.
Deviations of measured kinetic energies to true values can be expected because of scale errors in the analyzer hard- or software, because of parasitic (undesired) potentials emerging in conjunction with the liquid-jet passage through the glass nozzle, and because of the work-function difference between liquid jet and analyzer \cite{Pugini}.
We assume that the parasitic potentials are constant (but not necessarily identical) within each series of hypersatellite Auger and normal Auger calibration spectra, and within the series of calibration spectra used to determine the value of the KVV Auger energy.
We then proceeded as follows: from a series of liquid-water photoemission spectra below the single-ionization K-edge we firstly found the photon-energy calibration from the positions of the O~$1s$ photoemission peak measured with second-order light, and the 2a$_1$ inner-valence line or the vertical ionization-potential position (1b$_1$) measured with first-order light, all within the same scan (binding energies used: 538.1~eV, 31.07~eV, and 11.33~eV, respectively \cite{Thuermer2021}).
We then calibrated the kinetic-energy scale, and transferred this result to measurements of the KVV Auger spectrum at nominal photon energies of 600~eV and 640~eV.
Within our error estimate these gave the same result of 502.7(3)~eV for the position of the maximum.
We assume that the same value applies also at 1400~eV photon energy. 
This is supported by a large, unpublished data set yielding 502.6(2)~eV as the asymptotic value of the Auger energy towards high photon energies \cite{ThuermerPCI} and agrees reasonably with an older published value of 503~eV \cite{Winter2007a}.
\subsection{Ratio of double- to single-core-hole creation}
Under some reasonable assumptions, we can use the intensity ratio of the hypersatellite Auger to the normal liquid-phase Auger spectrum, from a singly ionized core, as an estimate for the relative probability of single-photon double-core-hole vs.\ single-core-hole creation.
Among the data sets recorded for this publication, we identified one for normal water and one for deuterated water that featured a statistically meaningful series of hypersatellite Auger data {\em and} a measurement of the normal Auger spectrum under identical conditions.
For these two data sets we proceeded as follows.

We determined the intensity of the double-core-hole decay as outlined above, but applied the background-subtraction procedure to each individual sweep of the data set (after selection by the normalized-cross-correlation criterion).
Intensity fluctuations of the individual sweeps (see Supplementary Fig.~\ref{sfig:traces}) were corrected to the average value of the subset of data measured temporally closest to the reference single Auger spectrum, on a sweep-by-sweep basis.
Integrating intensity from 540~eV to 575~eV of kinetic energy, this gave two sets of hypersatellite Auger intensities, each with a large scatter, but with meaningful relative errors around 15\% when the standard deviation of the mean was performed.

\begin{figure}[htb!]
\includegraphics{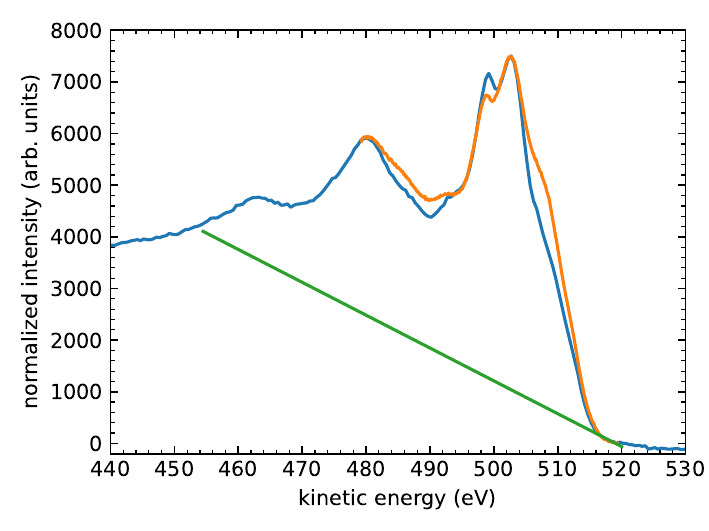}%
\caption{
Normal KVV Auger spectrum of liquid water (orange), recorded under the same conditions as the Auger hypersatellite spectra.
A spectrum showing the full energy range of Auger decay (blue) and a linear background applied to determine the Auger intensity (green) are also shown.
\label{sfig:area}}
\end{figure}

To determine the normal Auger intensity, we have to partition the measured spectrum into Auger and scattering contributions.
Additionally, the normal Auger spectra that were measured under conditions identical to the DCH Auger did not cover the full energy range of Auger decay of a single core hole (Supplementary Fig.~\ref{sfig:area}).
We have therefore taken a wide-range spectrum, similar to the one shown in Fig.~3(a) of the main text, and normalized the range pertinent to Auger decay of single core holes to the intensity scale of the reference single-core-hole Auger spectra.
We then determined from the wide-range spectrum a linear background to the Auger features  (Supplementary Fig.~\ref{sfig:area}).
Given our lack of information about the true background shape resulting from singly and multiply scattered photoelectrons, secondary electrons generated by impact ionization, and scattered Auger electrons, we consider this simple, parameter-free estimate of the background height to be appropriate for our purpose.
After background subtraction, the intensity of the normal Auger spectrum was determined in the kinetic-energy range [455.0,517.3]~eV \cite{Moddeman, Winter2007a}.
The same background was applied to the reference spectra measured under identical conditions to the DCH traces, and the area result was corrected for the contribution of the energy interval missing at the low-kinetic-energy end (see orange trace in Supplementary Fig.~\ref{sfig:area}).

From these two results the ratio of double-core-hole to single-core-hole Auger intensity was found as 1/626 and 1/777 for the normal and the deuterated water data, respectively.
We do not assign significance to the difference between the results for the two isotopes, but rather quote as the final result a value of 1/700 with an uncertainty of $\pm$33\%, comprised mostly of the fluctuations of the underlying hypersatellite data and systematic errors in peak-to-background separation.
Note that the resulting double-to-single ionization ratio of 0.0014(5) does not only contain the pure core-ionization processes, but also small contributions of ionization accompanied by additional valence shake processes.
\begin{figure}[htb!]
\includegraphics{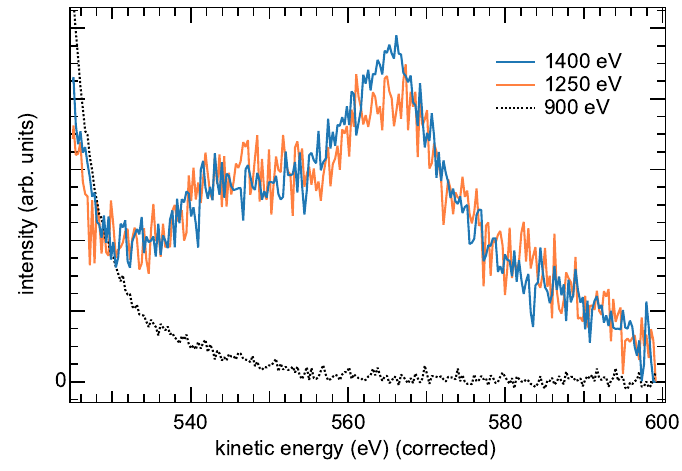}%
\caption{Photoemission spectrum of water recorded at various photon energies.
A linear background was removed from all spectra, and traces were scaled to aid in comparison.
The legend shows the excitation photon energy, with 900~eV below and 1250~eV as well as 1400~eV above the threshold for O~$1s^{-2}$ core-hole creation.
\label{sfig:hnu_c}}%
\end{figure}
\section{Reference spectra below the DCH threshold}
It is conceivable that other types of satellites are placed in the energy region in which we have located the Auger hypersatellites from double-core-hole decay. 
In Supplementary Fig.~\ref{sfig:hnu_c}, we, therefore, show the same energy region recorded with photon energies below (900~eV) and above (other photon energies) the threshold for double-core-hole excitation.
The signal level is generally lower in the 900~eV spectrum, as the scattering tail of the O~$1s$ main line at this photon energy is below the kinetic-energy interval that was scanned.
The figure shows that the features we assign to double-core-hole decays are absent in the 900~eV spectrum, but are present in the two spectra taken above the double-core-hole ionization potential.
The similarity of the two spectra recorded at different photon energies above the DCH threshold corroborates that the spectral feature we discuss corresponds to pure hypersatellite signal.
\section{Gas-to-liquid ratio}
In any liquid-jet photoemission experiment, a gas-phase signal from the condensing vapor surrounding the jet is present to some extent.
Its level can be assessed from spectral features that are separated in kinetic energy from their liquid counterparts and typically exhibit less spectral broadening.
In panel (a) of Supplementary Fig.~\ref{sfig:lqgas}, we show the normal KLL Auger spectrum of our liquid sample, with the position of the sharp, most intense peak of the gas-phase Auger spectrum \cite{Moddeman} marked by an arrow.
The amount of gas-phase signal present is low in a typical data set, but notably higher in our first measurements, due to the worse vertical focusing of the synchrotron-radiation beam in this campaign.
Occasionally we have also recorded O~$1s$ photoelectron spectra of the liquid sample, leaving all conditions unchanged.
These allow for a better quantitative assessment of the gas-phase contribution, placing the typical fraction of electrons from the gas phase at 10\% or lower [Supplementary Fig.~\ref{sfig:lqgas}(b)].
For spectra acquired during the first campaign in 2018 this fraction might have been somewhat higher, as the vertical demagnification of the P04 photon spot was not yet used.
Based on these results we estimate that the signal from the hypersatellite Auger spectrum coming from gas-phase molecules does not have a sizeable effect in the liquid-phase spectra we discuss, but refer the reader back to the main text for a discussion.
\section{Search for photoionization satellite lines}
Within the course of this project we have also searched for satellite lines originating from normal photoionization of an O~$1s$ electron going along with an excitation of the other O~$1s$ electron at the same site to an unoccupied valence state.
These processes were seen in the photoemission spectra of gaseous water, as discussed most recently in \cite{Marchenko2020} and elsewhere \cite{Mucke2013, Carniato2015}.
Since the overall charge of the photoexcited state is $+1$, discrete photoemission lines can be expected in a conventional photoionization experiment.
Their binding energy can be below or above the DCH binding energy \cite{Tenorio2021}, found at 1171~eV for gas-phase water \cite{Mucke2013}.
We have not observed any such feature in the binding-energy interval [1140,1210]~eV at a photon energy of 1400~eV, within the sensitivity of our experiment.
This sensitivity can be assessed from an observation of the Na~$1s$ core-level photoemission line of the 50~mM NaCl admixture to our sample solution, at its (as-measured) binding energy of 1076.5~eV and with a width of 1.5~eV.
To compare a hypothetical DCH satellite line to this observed feature, we need to compare their respective cross sections.
For this comparison we used calculated single-photoionization cross sections (Ref.~\cite{Yeh}), and, for O~$1s$, scaled them down by the Auger hypersatellite to conventional KVV Auger area ratio (see above) and by an additional factor of ten, estimated for the production ratio of all double-core-hole states represented by the Auger hypersatellites to the O~$1s$ singly ionized DCH satellite states \cite{Goldsztejn2017}.
For Na~$1s$ we took into account the dilution of NaCl compared to our water sample.
As a result of this exercise, we estimate that an improvement of sensitivity in our experiment by a factor of about five would be needed to spectroscopically observe an O~$1s$ DCH satellite line from liquid water, if it were confined to a similar width of kinetic energy than the Na~$1s$ band.
If in liquid water the satellites appear over a wider energy range because of interactions in the liquid, an even higher improvement in sensitivity would be needed.

%

\section{Further calculation results}

\begin{figure}[htb!]
\includegraphics{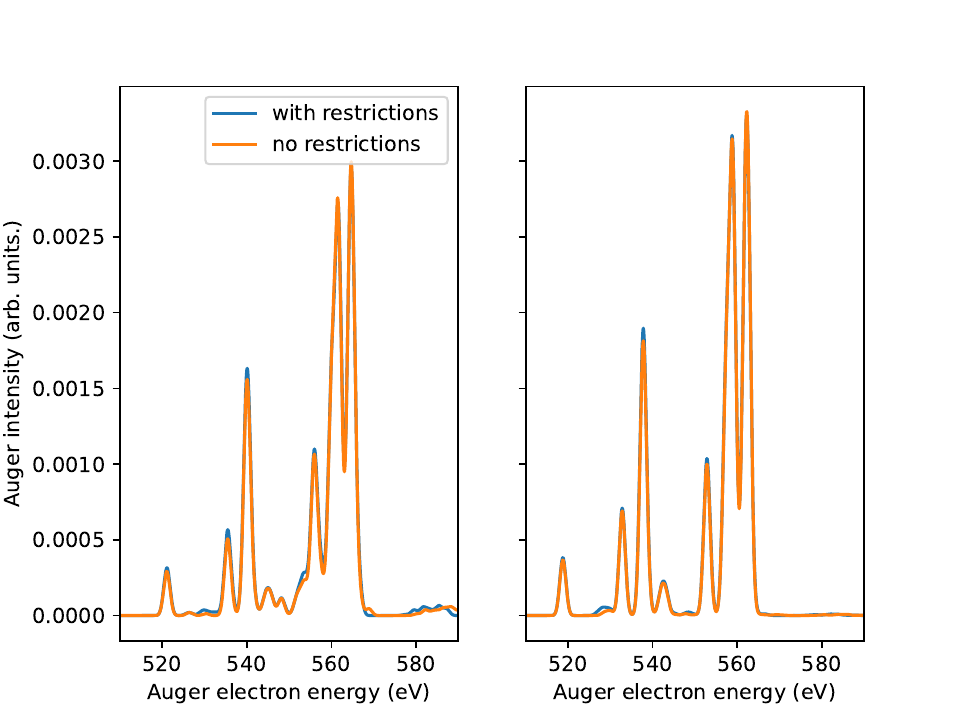}\\
\caption{
Calculated KK-VV Auger spectrum for the water dimer.
For the left spectrum the initial double core hole is located on the hydrogen-bond donor molecule, for the right spectrum it is located on the hydrogen-bond acceptor molecule.
Two results are shown: with configuration restrictions and without (see text for further details).
\label{sfig:auger_dimer}}%
\end{figure}

To make the computation of the Auger spectra feasible, we employed restriction on the configurations for the CI expansion.
This restriction is tested here for a water dimer, where the double core hole is either located on the hydrogen-bond-accepting molecule (acceptor) or on the hydrogen-bond-donating molecule (donor).
For both scenarios, calculated Auger spectra employing a fixed geometry as well as a convolution with a Gaussian function of 1.7~eV width (full width at half maximum) are shown in Supplementary Fig.~\ref{sfig:auger_dimer}.
The blue curve shows the results obtained with the configuration restriction, that is, for the configuration interaction only those configurations are considered where maximal one valence hole is not located on the respective core-ionized water molecule.
This restriction reduces the number of considered spin configurations from 14,960 to 7,496, whereas for the water pentamer this reduction is much larger.
As one can see, it has almost no effect on the resulting Auger spectrum.
The largest differences can be seen in the small intermolecular Auger contributions that appear around 580~eV.

Remarkably, considerable differences appear between the donor and the acceptor Auger spectra.
The former is considerably shifted to higher energies.
Whereas ICD-like contributions for the donor are visible, they are almost absent for the acceptor.
This highlights that the different role in hydrogen bonding has considerable effects on the DCH Auger spectrum.

\newpage
\section*{References}

The full set of experimental data has been deposited to Ref.~\cite{dataset}.

\bibliography{supplementary}